\documentclass[aps,superscriptaddress,showkeys]{revtex4}

\usepackage[dvips]{graphicx}

\begin{document}

\title{Compact stars made of fermionic dark matter}

\author{Gaurav Narain}
\email{gauhari@iitk.ac.in}

\affiliation{
Indian Institute of Technology, 
Kanpur, Uttar Pradesh, 
India}

\affiliation{
Institut f\"ur Theoretische Physik, 
J. W. Goethe Universit\"at,
Max von Laue-Stra\ss{}e 1,
D-60438 Frankfurt am Main,
Germany}

\author{J\"urgen Schaffner-Bielich}
\email{schaffner@astro.uni-frankfurt.de}

\affiliation{
Institut f\"ur Theoretische Physik, 
J. W. Goethe Universit\"at,
Max von Laue-Stra\ss{}e 1,
D-60438 Frankfurt am Main,
Germany}

\author{Igor N. Mishustin}
\email{mishustin@fias.uni-frankfurt.de}

\affiliation{
Frankfurt Institute for Advanced Studies,
J. W. Goethe Universit\"at,
Max von Laue-Stra\ss{}e 1,
D-60438 Frankfurt am Main,
Germany}

\affiliation{
The Kurchatov Institute,
Russian Research Center,
123182 Moscow, Russia}

\date{\today}

\begin{abstract}
  Compact stars consisting of fermions with arbitrary masses and
  interaction strengths are studied by solving the structure equation of
  general relativity, the Tolman-Oppenheimer-Volkoff equations. Scaling
  solutions are derived for a free and an interacting Fermi gas and
  tested by numerical calculations. We demonstrate that there is a
  unique mass-radius relation for compact stars made of free fermions
  which is independent of the fermion mass. For sufficiently strong
  interactions, the maximum stable mass of compact stars and its radius
  are controlled by the parameter of the interaction, both increasing
  linearly with the interaction strength. The mass-radius relation for
  compact stars made of strongly interacting fermions shows that the
  radius remains approximately constant for a wide range of compact star
  masses.
\end{abstract}


\maketitle

\section{Introduction}

Compact stars, white dwarfs and neutron stars, are one of the most
enigmatic astrophysical objects (for an introduction to the physics of
compact stars we refer to the excellent textbooks
\cite{Shapiro_book,Glen_book,Weber_book} and to recent pedagogical
papers \cite{Balian99,Silbar04,Jackson05,Macher05,Sagert05}). The first
successful description of compact stars was proposed by Fowler in 1926
\cite{Fowler26}, who first realized that Fermi-Dirac statistics is
responsible for the high degeneracy pressure which holds up the star
against gravitational collapse. Shortly afterwards Chandrasekhar applied
effects of special relativity to the equation of state (EoS) of a
degenerate Fermi gas and demonstrated the existence of a maximum mass
for such stars beyond which they are unstable against collapse: the
famous Chandrasekhar mass limit, $M_{ch}\approx 1.4 M_\odot$
\cite{chandra31}.

In 1932 Chadwick discovered the neutron and Heisenberg presented his
theory of isospin for nucleons suggesting that the neutron is a fermion
with spin-$\frac{1}{2}$ like the electron. The ideas of Fowler and
Chandrasekhar were then immediately extended to the case of degenerate
neutrons and a new form of compact stars, neutron stars, was predicted
by Landau \cite{Landau32}.  The first numerical calculations for a
neutron star within general relativity were performed by Oppenheimer and
Volkoff in 1939 \cite{OV39}. They computed a maximum stable mass of $
0.75 M_{\odot} $ for a free Fermi gas of neutrons beyond which the star
would be unstable and collapse into a black hole. The maximum mass limit
for neutron stars is now well known as the Oppenheimer-Volkoff mass
limit. Contrary to the case of the electron gas in white dwarfs, there
are sizable contributions to the mass limit for neutron stars due to
strong interactions between the neutrons. These interactions can be well
modelled by an effective repulsive potential which can increase the
maximum stable mass of a neutron star to about $3 M_{\odot}$. The
presence of hyperons in the core of neutron stars leads to a softening
of the equation of state and a reduction of the maximum mass
\cite{Ambart60,Glen85,Knorren95a,Knorren95b,SM96}.

As of today, new massive fermions are known within the standard model
and many more have been predicted, in particular also as candidates for
dark matter. In the year 1964, Gell-Mann and Zweig proposed the idea of
quarks, putting forward the notion that neutrons and protons are
composed of quarks. Ivanenko and Kurdgelaidze \cite{Ivan65} studied a
quark core in massive compact stars and Itoh \cite{Itoh70} calculated
the mass-radius relation of a quark star. If a compact star consists of
quarks only, including besides the light quarks also the strange quark,
they are dubbed strange stars \cite{Witten84,Haensel86,Alcock86}. Those
exotic objects might be bound by strong interactions only, contrary to
ordinary neutron stars and white dwarfs which are bound by gravity. The
physics of compact stars with a quark core and strange stars is
now a field of active research (for a recent review we refer to
\cite{Weber05}) and has found its place in modern textbooks
\cite{Glen_book,Weber_book}.  Besides quarks, other fermions in the form
of massive neutrinos are now well known to exist in nature.  New types
of fermions predicted in extensions of the standard model include the
supersymmetric particles, the neutralino, the gravitino, and the axino,
which are also candidates for dark matter (for a review see
e.g.~\cite{Baltz04}).

Now the idea put forward by Fowler many years ago could be used to
speculate on compact stars made out of exotic fermions, stabilised by
the degeneracy pressure in the same way as in the case of neutron stars
and white dwarfs. The present investigation is dealing with the
following questions: What is the maximum stable mass of compact stars as
a function of the fermion mass at zero temperature? What will happen if
a repulsive interaction is included in the equation of this fermionic
matter?

The paper is organised as follows: in section \ref{sec:derivationTOV} we
briefly recapitulate the structure equation for compact stars, the
Tolman-Oppenheimer-Volkoff (TOV) equation. In section
\ref{sec:scalingTOV}, we discuss general scaling solutions for compact
stars with an arbitrary equation of state, in particular for a free
Fermi gas and for an interacting Fermi gas. We show that Landau's
estimate for the maximum mass and the corresponding radius is an exact
scaling solution of the TOV equation for a free Fermi gas.  Section
\ref{sec:numericalTOV} is devoted to our numerical solution of the TOV
equations.  The equation of state for interacting fermions as well as
extended scaling solutions of the TOV equations are discussed.  We show
that there is one unique mass-radius relation for compact stars made of
free fermions if it is properly rescaled by the Landau mass and the
Landau radius. We also find that for strongly interacting fermions the
maximum mass and its radius are controlled by the interaction terms.
The mass-radius relation changes such that there is a constant radius
for a wide range of masses. Finally, in section \ref{sec:summary} we
summarise and discuss our findings.

\section{The structure equation for compact stars}
\label{sec:derivationTOV}

Throughout this paper we will be dealing with models of compact stars
where effects of general relativity are included for a consistent
description as in the case of ordinary neutron stars.  The typical mass
of a neutron star is of the order of $\sim 1 M_{\odot}$ with typical
radii of about 10 km, i.e.\ $10^{-5} R_{\odot}$. Hence, the
gravitational potential on the surface of neutron star will be $10^{5}$
times stronger compared to that of the Sun. Moreover, the corresponding
Schwarzschild radius $r_s=2GM/c^2$ is about 3 km in this case.  Under
such conditions, the curvature of space-time can not be ignored and
general relativity is needed to describe the structure of such compact
objects.

In order to find the structure of space-time created by the presence of
a compact star one needs to solve the Einstein's field equations. For
simplicity we assume that the metric is spherically symmetric and
static, i.e.\ the Schwarzschild metric. In addition, the energy momentum
tensor is assumed to be that of an ideal fluid,
\begin{equation}
T_{\mu \nu} = pg_{\mu \nu} + (p+\rho)U_{\mu}U_{\nu} \quad ,
\end{equation}
where $p$ and $\rho$ denote the pressure and energy density, and
$U_{\mu}$ the four velocity of the fluid. Using the Einstein's field
equation and the condition for hydrostatic equilibrium, $U_\mu=(1,0)$,
one arrives at the following equations describing the structure of a
compact star
\begin{eqnarray}
\label{tov1}
\frac{dp}{dr} &=& -\frac{GM \rho}{r^2} \left(1 +
  \frac{p}{\rho}\right)\left(1 + \frac{4 
  \pi r^3 p}{M}\right)\left(1 - \frac{2GM}{r}\right)^{-1} \quad ,\\ 
\label{tov2}
\frac{dM}{dr} &=& 4 \pi r^2 \rho \quad ,
\end{eqnarray}
which are just the Tolman-Oppenheimer-Volkoff (TOV) equations
\cite{Tolman34,Tolman39,OV39} (note, that throughout the paper, we are
using natural units by setting $\hbar=c=1$).  The detailed derivation of
the equation (\ref{tov1}) can be found in standard textbooks (see e.g.\ 
\cite{Weinberg72,Misner73,Glen_book,Weber_book}). This is the same
equation of hydrostatic equilibrium as in the case of Newtonian gravity
just modified by three correction factors (in the r.h.s.) due to effects
of general relativity. The equation (\ref{tov2}) simply defines the
quantity $M(r)$, the amount of energy contained within the radius $r$.

The unknown functions in eq.~(\ref{tov1}) and eq.~(\ref{tov2}) are
$\rho(r)$, $p(r)$ and $M(r)$. For a given equation of state, relating $p$
and $\rho$, appropriate initial and boundary conditions are needed to
solve the above set of equations.  The radius of the star, $R$, is found
by using the condition that the pressure vanishes at the surface of
star.  The mass $M(0)$ must be zero at $r=0$ and $M(R)$ gives the total
mass of the star at $r=R$. The central pressure is calculated from the
equation of state once the central energy density $\rho(0)=\rho_0$ is
given as the initial condition.

\section{Scaling the TOV equation}
\label{sec:scalingTOV}

It is easy to see, that the TOV equation contains two essential
dimensional quantities, $G$ which can be conveniently expressed in terms
of the Planck mass as $G=M_p^{-2}$, and the fermion mass $m_f$ which
characterises the equation of state. In this section, we show how the
TOV equation can be transformed to scale-independent variables composed
of $M_p$ and $m_f$. There are several reasons for such an approach. One
reason is that the computational treatment of differential equations
benefits from a dimensionless format.  The other reason is that a scaled
equation needs to be solved only once.  As soon as the general solution
is found one can just rescale it by appropriate (dimensionful) factors
to get the result for specific (astro-)physical cases. The TOV equation
can be scaled in the following ways.

\subsection{Landau's argument for deriving the maximum mass of compact stars}

Landau presented a very elegant argument for deriving the maximum
mass of a compact star \cite{Landau32} (for details see also the
treatises in \cite{Harrison_book,Shapiro_book,Straumann_book}). He used
only Newtonian gravity, special relativity and Fermi-Dirac statistics to
estimate the maximum mass and the corresponding radius of a compact
star.

For a star made of free fermions at zero temperature the Fermi momentum
$k_{F}$ is related to number density $n$ by the following relation
(below we suppress the dependence on the statistical degeneracy factor
assuming $g=2$)
\begin{equation}
n = \frac{k_{F}^3}{3 \pi ^2} = \frac{N}{4\pi/3 R^3} \quad ,
\label{num:den}
\end{equation}
where $N$ is the total number of fermions in a star. For simplicity,
here we consider a star of uniform number density and of radius $R$.
Solving for $k_F$, one gets
\begin{equation}
k_F = \left(\frac{9 \pi}{4}\right)^{1/3} \frac{N^{1/3}}{R} 
\quad .
\end{equation}
The total mass of the star is given solely by the vacuum fermion mass,
$m_{f}$, as
\begin{equation}
M = m_{f} N \quad .
\end{equation}
Now, let us consider a fermion on the surface of the star. Its energy
is given by
\begin{equation}
E(R)  = -\frac{GM m_{f}}{R} + \left(\frac{9 \pi}{4}\right)^{1/3}
\frac{N^{1/3}}{R} \quad ,
\end{equation}
where the first term gives the gravitational energy and the second term
comes from the kinetic energy of an ultrarelativistic fermion on the top
of the Fermi distribution. For small values of $R$ and negative energy
$E$, the gravitational attraction overcomes the degeneracy pressure
causing a collapse.  For positive $E$ the degeneracy pressure exceeds
the gravitational attraction and the star will expand until the particle
density drops so much that $k_F\sim m_f$. For the non-relativistic gas
the kinetic energy per particle is $3k_F^2/(10 m_f)$, i.e.\ it
changes with the radius as $R^{-2}$. This means that gravitation
will finally prevail and the expansion will be terminated. As a result,
a stable minimum will develop in $E(R)$.

A good estimate of the maximum possible number of fermions in a compact
star can be obtained by considering the limiting case $E=0$, when the
gravitational energy is exactly equal to the Fermi energy of the
degenerate Fermi gas. Then we can express the maximum number of fermions
as 
\begin{equation}
N_{max} = \left(\frac{9 \pi}{4}\right)^{1/2} \frac{M_{p}^3}{m_{f}^3} 
\quad .
\end{equation}
The maximum mass of the star is obtained from the relation $M = m_f N$,
hence
\begin{equation}
M_{max} \sim \frac{M_{p}^3}{m_{f}^2} \quad .
\end{equation}
An estimate for the corresponding radius of the maximum mass star can be
obtained by assuming that the kinetic energy of the fermion on the
surface is equal to its mass, i.e.\ at the border of becoming
relativistic, $k_F\approx m_f$, which gives for the minimum radius of the star
\begin{equation}
R_{min} \sim \frac{M_{p}}{m_{f}^2} \quad .
\end{equation}
For a neutron star, with the fermion mass taken to be that of the
neutron $m_n \approx 1$~GeV, the above relations give $M_{max} \approx
1.63 M_{\odot} $ and $R_{min} \approx 2.41$ km which is a reasonable
estimate.  Note, that Landau's argument can be well applied also for
white dwarfs.  Here, one has to take care of the fact that the mass of
the white dwarf is determined by the nucleon mass while the degeneracy
pressure is provided by the electrons. Therefore, the maximum mass for
white dwarfs, the Chandrasekhar mass limit, turns out to be similar to
the one for neutron stars, while the radius increases by the ratio of
the nucleon and the electron masses, i.e.\ by about a factor 2000, to
about 4000 km (all these values have to be corrected for the charge to
mass ratio of nuclei, see e.g.\ \cite{Straumann_book}).

\subsection{The equation of state in dimensionless form}

The equation of state for a free gas of fermions at zero temperature
$p(\rho)$ can be calculated via explicit expressions for the energy
density and pressure:
\begin{eqnarray}
\rho  &=& \frac{1}{\pi ^2} \int_0^{k_F} k^2 \sqrt{m_f^2+k^2} dk  
\nonumber\\
&=& \frac{m_{f}^4}{8 \pi ^2}\left[ (2 z^3 + z)(1+ z^2)^{\frac{1}{2}} - \sinh
^{-1}(z) \right] \equiv m^4_f \rho' 
\label{eqf:den1}
\\
p &=& \frac{1}{3 \pi^2} \int_0^{k_F} \frac{k^4}{ \sqrt{m_f^2+k^2}} dk 
\nonumber\\
&=& \frac{m_{f}^4}{24 \pi ^2}\left[ (2 z^3 - 3 z)(1+ z^2)^{\frac{1}{2}} + 3
\sinh ^{-1}(z) \right] \equiv m^4_f p'  
\label{eqf:pres1}
\end{eqnarray}
using natural units of $m_f^4$ and defining the relativity parameter $z
= k_F/m_f$.

\begin{figure}
\includegraphics[angle=-90,scale=.45]{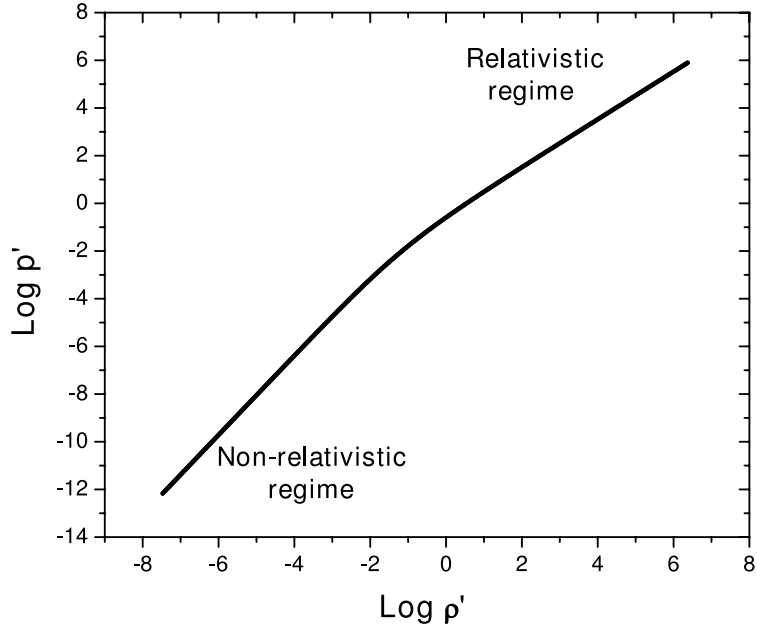}
\caption{The equation of state, the dimensionless pressure $p'$ versus
  the dimensionless energy density $\rho'$ for a free Fermi gas at zero
  temperature. The two limits, the non-relativistic one for $z\ll 1$ and
  the ultra-relativistic one for $z\gg 1$, can be clearly seen.}
\label{fig:eqf}
\end{figure}

We introduce now the following dimensionless quantities for the mass and
the radius of the star, 
\begin{equation}
M' = \frac{M}{M_L} \quad \mbox{with} \quad
M_L = \frac{M_{p}^3}{m_{f}^2} 
\qquad \mbox{and} \qquad
r' = \frac{r}{R_L} \quad \mbox{with} \quad
R_L = \frac{M_{p}}{m_{f}^2}
\end{equation}
where $M_L$ and $R_L$ denote the maximum mass and corresponding radius
as given by Landau's arguments (note that $R_L$ is equal to half the
Schwarzschild radius). Using the above definitions and substituting them
into eq.~(\ref{tov1}) and eq.~(\ref{tov2}) with $G = M_p^{-2}$, one
obtains the following dimensionless form of the TOV equations:
\begin{eqnarray}
\frac{dp'}{dr'} &=& -\frac{M' \rho '}{r'^2} \left(1 + \frac{p'}{\rho
    '}\right)\left(1 + 
\frac{4 \pi r'^3 p'}{M'}\right)\left(1 - \frac{2M'}{r'}\right)^{-1} 
\label{nd:tov1} \\
\frac{dM'}{dr'} &=& 4 \pi r'^2 \rho '
\label{nd:tov2}
\end{eqnarray}
In general, the equation of state can not be expressed in a simple
polytropic form $p' \propto {\rho'}^\gamma$ with a constant $\gamma$.
However, this can be done for two limits of the relativity parameter
$z$. In the non-relativistic limit, $z \ll 1$, we get a polytropic law
with $\gamma=5/3$, i.e.
\begin{equation}
p' \propto {\rho'} ^{5/3} \qquad (z\ll 1) \quad .
\label{nonrel}
\end{equation}
In the ultra-relativistic limit, $z \gg 1$, the equation of state
approaches a polytrope of $\gamma =1$, i.e.
\begin{equation}
p' = \frac{\rho'}{3} \qquad (z\gg 1) \quad .
\label{ultra:rel}
\end{equation}
As the equation of state is a function of the relativity parameter $z$,
one can compute it parametrically in a tabular form for a desirable
interval of $z$.  Fig.~\ref{fig:eqf} depicts the resulting dimensionless
pressure versus the dimensionless energy density in a double logarithmic
plot. The curve exhibits two different slopes for small and large values
of $\rho'$. The larger slope at small values of $\rho'$ shows the
non-relativistic regime, the smaller slope at larger values of $\rho'$
the relativistic regime as expected.

For the case of white dwarfs, the following changes have to be made to
the equation of state, eq.~(\ref{eqf:den1}) and eq.~(\ref{eqf:pres1}):
\begin{enumerate} 
\item As the pressure is given by the degeneracy pressure of electrons,
  the expression for the pressure remains the same where $m$ is the mass
  of the electron.
\item The energy density is a sum of two terms, one coming from the mass
  density of protons and neutrons while the other from the kinetic
  energy of electrons.
\end{enumerate}
For a white dwarf consisting of nuclei with the atomic mass $A$ and
atomic number $Z$ the mass density of protons and neutrons reads (small
effects from the nuclear binding energy are disregarded here)
\begin{equation}
\rho_{1} = n \cdot m_{N} \cdot \frac{A}{Z}
\label{wd:massden}
\end{equation}
where due to charge neutrality the number density of protons $n$ is
equal to the number density of electrons which is given by
eq.~(\ref{num:den}). The total energy density is now written as
\begin{equation} 
\rho = \rho_{1} + \frac{m_{e}^4}{\pi ^2} \int_0^{z} x^2 \sqrt{1+x^2} dx
\quad .
\end{equation}
The second term here gives the relativistic energy density of electrons
(note that here $z=k_F/m_e$). The resulting equation of state can be
scaled in the same way as done before. Defining the dimensionless pressure
and dimensionless energy density as $ p' = p/m_{e}^4$ and $\rho' =
\rho/m_{e}^4$, we can rewrite the above equation in the dimensionless form
\begin{equation} 
\rho' = \frac{z^3}{3 \pi^2}  \frac{m_{N}}{m_{e}} \frac{A}{Z} +
\frac{1}{\pi ^2} \int_0^{z} x^2 \sqrt{1+x^2} dx \quad .
\end{equation}
The pressure $p'$ is given by eq.~(\ref{eqf:pres1}) where $m_{f} =
m_{e}$. In the non-relativistic case, $z \ll 1$, the equation of state
simplifies to eq.~(\ref{nonrel}) . On the other hand, in the
relativistic limit $1\ll z \ll 6\cdot 10^3$, when the first term on the
r.h.s.\ dominates, the equation of state for
white dwarfs becomes
\begin{equation}
p' \propto {\rho'} ^{4/3} \quad .
\end{equation}
The proportionality constant in both limits, non-relativistic and
relativistic ones, depends on $A$ and $Z$. Note, that the special case
for white dwarfs emerges due to the presence of oppositely charged
fermions which are neutralising each other. The heavier one determines
the energy density, while the lighter one the pressure. For a pure
charge-neutral Fermi gas, as we will consider later, the
ultra-relativistic limit will be realized, as for the case of neutron
stars.

\subsection{Scaling solution for self-bound compact stars}

Another scaling solution is well known \cite{Witten84,Haensel86} for the
special form of the equation of state (see also \cite{Glen_book} and
references therein):
\begin{equation}
p = s( \rho - \rho_0)
\label{eq:mit}
\end{equation}
where $s$ and $\rho_0$ are constants. Note, that this equation of state
has a vanishing pressure at a finite energy density $\rho=\rho_0$. This
property allows for the existence of self-bound balls of any size,
not necessarily of astronomical scale. These balls are stabilised by
other interactions not by gravity as in the case of neutron stars.

The scaling relations $p' = p/\rho_0$, $\rho ' = \rho/\rho_0$, $r' =
\sqrt{\rho_0} (r/M_{P})$ and $M' = \sqrt{\rho_0} (M/M_{P}^3)$ change
the TOV equation to a dimensionless form.  One can then solve this
equation numerically. If the mass and the radius are
known for some particular value of $\rho_0$, then for some other value
$\rho_0'$ the radius and mass will be $R(\rho_0') =
\sqrt{\rho_0'/\rho_0} R(\rho_0 )$ and $M(\rho_0') =
\sqrt{\rho_0'/\rho_0} M(\rho_0)$, respectively. Hence, both the mass and
the radius scale with $1/\sqrt{\rho_0}$.

The above equation of state is actually the one of the MIT bag model
often used for describing cold and massless (strange) quark matter. The
corresponding self-bound compact stars are dubbed strange stars
\cite{Witten84,Haensel86,Alcock86}. Most studies of quark stars and
strange stars utilise the equation of state derived from the MIT bag
model.
Ignoring effects from a finite quark
mass, this model gives the equation of state
\begin{equation}
p = \frac{1}{3} \left( \rho - 4B \right) \quad ,
\label{eq:mitbag}
\end{equation}
where $B$ is the bag constant.  Interestingly, the equation of state for
an interacting cold gas of massless quarks within perturbative quantum
chromodynamics can be approximated by the same form of the equation of
state \cite{FPS01}.  Standard values for the MIT bag constant are around
$B^{1/4} = 145$ MeV as follows from fits to hadron masses, which results
in maximum masses of about $2.0 M_\odot$ at a radius of about 11 km
\cite{Baym76,Witten84,Haensel86}, which are actually very close to the
ones of realistic neutron star models. It is worth noting, that these
values of $B$ are obtained in fits including the chromomagnetic
interaction. In simplified versions of this model, disregarding the
interaction effects, higher values, $B^{1/4} \simeq 200$ MeV, are needed
to preserve the stability of normal nuclear matter \cite{Mishustin02}.

\subsection{Scaling in a general case}

Consider the TOV equation for the pressure, eq.~(\ref{tov1}). We make
the observation that the three correction factors from general
relativity are already in a dimensionless form.  Therefore, from the
first factor $(1+p/\rho)$ one can scale pressure and energy density by a
common factor $\epsilon _{\circ}$ as $ p' = p/\epsilon_{\circ}$ and
$\rho ' = \rho/\epsilon_{\circ}$, respectively. Similarly, one defines a
dimensionless mass and radial coordinate via $M' = M/a$ and $r' = r/b$.
Plugging these definitions into the second factor $(1 + 4 \pi r^3 p/M)$
one arrives at a dimensionless number $b^3\epsilon_{\circ}/a$ which we
equate to one.  Similar reasoning for the third factor $(1 -
2GM/r)^{-1}$ results in $a/(M_{p}^2 b)$ as the dimensionless number
which again is equated to one. So we get finally the following scaling
conditions
\begin{equation}
\frac{b^3 \epsilon_{\circ}}{a} = 1
\qquad \mbox{ and } \qquad 
\frac{a}{M_{p}^2 b} = 1 \quad .
\end{equation}
Solving for $a$ and $b$ from the above equations one gets the following
expressions
\begin{equation}
a = \frac{M_{p}^3}{\sqrt{\epsilon_{\circ}}}
\qquad \mbox{ and } \qquad 
b = \frac{M_{p}}{\sqrt{\epsilon_{\circ}}} \quad .
\label{eq:scalefactors}
\end{equation} 
For a free gas of fermions at zero temperature, we know from the direct
calculation that $ \epsilon_{\circ} = m_{f}^4 $ (see eq.~\ref{eqf:den1}).
Substituting this value of $ \epsilon_{\circ}$ into the above equations
one finds $a = M_{p}^3/m_{f}^2$ and $b = M_{p}/m_{f}^2$ which are
exactly the same scaling factors as used originally by Landau. In the
case of the MIT bag equation of state, the scaling factors are given by
eqs.~(\ref{eq:scalefactors}) with $\epsilon_{\circ} = \rho_0 = 4B$.

We note in passing that according to Landau's argument it is sufficient
to incorporate special relativity and Newtonian gravity to obtain the
maximum mass $M_{max}$ and the corresponding minimum radius $R_{min}$ of
a cold compact star made of fermions. Interestingly, the same
dimensional forms for the maximum mass and minimum radius are found when
using dimensional reasoning applied to the full TOV equation of general
relativity.

\section{Numerical solution of the TOV equation}
\label{sec:numericalTOV}

\subsection{Stars made of free fermions}

For a free gas of fermions at zero temperature, the equations to be
solved are the dimensionless TOV eq.~(\ref{nd:tov1}) and
eq.~(\ref{nd:tov2}) along with the dimensionless equation of state given
in parametric form by equations (\ref{eqf:den1}) and (\ref{eqf:pres1}).
One starts with the dimensionless central pressure or energy density,
then solves the TOV equations from the center of the star to the surface
where the pressure becomes zero. The corresponding radial distance
defines the radius of the compact star.

\begin{figure}
\begin{center} 
  \includegraphics[angle=-90,scale=.45]{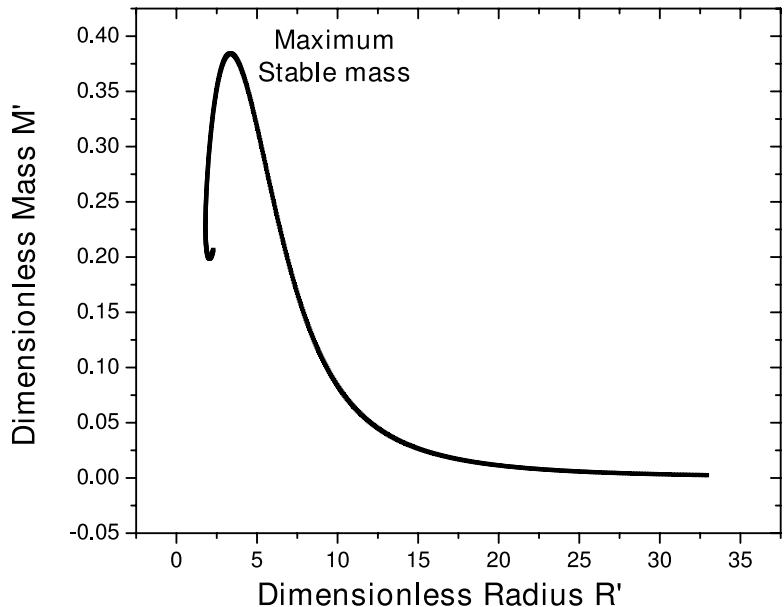}
\caption{Mass-radius relation in Landau units obtained by solving
  numerically the dimensionless TOV equation for a range of
  dimensionless central densities from $\rho' = 10^{-8} $ to $\rho' =
  10$. Note, that the curve does not depend on the mass of the fermions
  forming the compact star.}
\label{fig:MRrel}
\end{center}
\end{figure}

Since the dimensionless equation of state is given as a function of
parameter $z$, both quantities $p'$ and $\rho'$ are calculated for various
values of $z$ and expressed in a tabular form. To calculate $p'$ for a
given $\rho'$ and vice-versa, we use a simple linear interpolation. The
dimensionless TOV equation is solved using a fourth-order Runge-Kutta
algorithm. To plot the $M'$ versus $R'$ curve, $M'$ and $R'$ are
calculated for various dimensionless central densities.  The step size
of the dimensionless radius lies between $0.01$ (for very low
dimensionless central density) and $0.001$ (for very high
dimensionless central density) to have at least 2000 points for one
star configuration. The step size in the dimensionless central density
is adjusted to have 500 points between $10^{-8}$ and $10$.  The final
dimensionless mass-radius relation is plotted in Fig.~\ref{fig:MRrel}.

From the curve in Fig.~\ref{fig:MRrel} one notes that for a small
dimensionless mass the dimensionless radius is large. This behaviour
occurs for very small dimensionless central densities. The gravitational
attraction is small in this case making the dimensionless radius
large. As the dimensionless central density increases, the mass of the
star increases, too.  This leads to a stronger inward gravitational pull
and smaller dimensionless radii. Thus, as we increase the dimensionless
central density the mass increases while the radius decreases. A maximum
mass is reached for the dimensionless mass $M'_{max}=0.384$ at the
dimensionless radius $R'_{min}=3.367$.  The presence of the maximum in
the curve is generic and expected for an arbitrary fermion mass.  The
reason is that the energy density which generates the gravitational pull
inwards has to be balanced by the outward fermionic pressure. However,
the rate of change of the pressure with energy density is related to the
speed of sound which is bounded by the speed of light. This speed limit
puts a bound on the pressure increment with respect to changes in energy
density. Thus an increase of the central energy density results in
an increased gravitational attraction which cannot be compensated by
the corresponding additional pressure, that leads eventually to a
maximum mass limit \cite{chandra31,Landau32}.  It is easy to see
\cite{Shapiro_book}, that star configurations on the right hand side of
the maximum in Fig.~\ref{fig:MRrel} are stable whereas those on the left
hand side are unstable. The maximum in the curve proves the fact that
there exists a maximum stable mass for a fermion star, beyond which the
stars are unstable and collapse.

\begin{figure}
\begin{center} 
  \includegraphics[angle=-90,scale=.45]{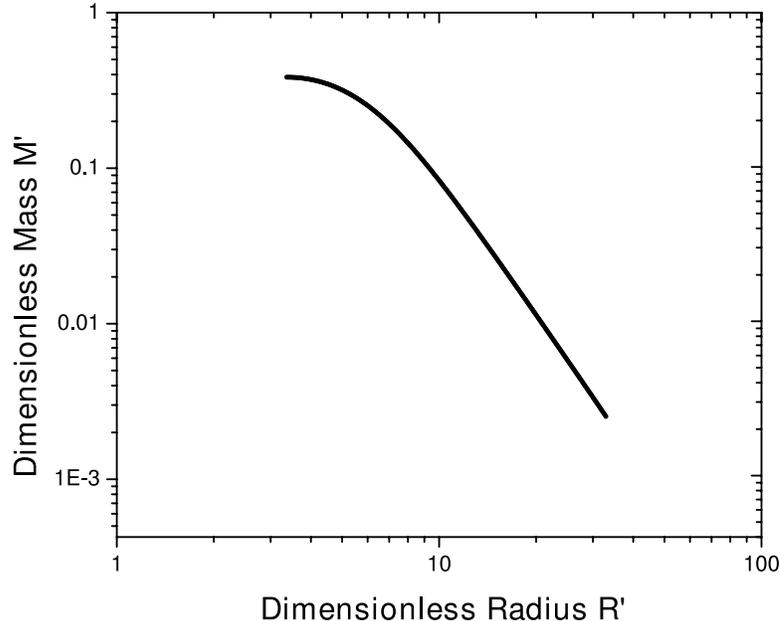}
\caption{The above graph shows the $M'$ versus $R'$ curve on a
  logarithmic scale for large radii.}
\label{fig:LogMR}
\end{center}
\end{figure}

Fig.~\ref{fig:LogMR} shows $M'$ versus $R'$ on a double-log scale.  For
large values of the dimensionless radius $R'$, the mass $M'$ and the
radius $R'$ follow the relation
\begin{equation}
M' \cdot {R'}^{\beta} = constant
\end{equation}
A linear fit to the curve for large $R'$ gives $ \beta
= 3.0001 \approx 3 $ and the constant is $90.97 \approx 91$. Putting these
values in the above relation we get
\begin{equation}
M' \cdot {R'}^{3} \approx 91
\label{eq:MR3rel}
\end{equation}
The above relation can be also derived analytically by considering a
non-relativistic degenerate gas of free fermions and Newtonian gravity.
Compact star configurations with a polytrope form of the equation of
state, $p\sim \rho^\gamma$, are given by the Lane-Emden function
\cite{Weinberg72,PadmanabhanII}. A non-relativistic gas of fermions has
a polytrope of $\gamma = 5/3$ and obeys the Lame-Emden equation of index
$(\gamma -1 )^{-1} = 3/2$. For a general polytrope, the relation between
$M'$, $R'$ and the central density $\rho'(0)$ are given by
\begin{equation}
M'\propto \rho'(0)^{(3\gamma - 4)/2} \quad \mbox{ and } \quad
R'\propto \rho'(0)^{(\gamma - 2)/2} \quad \Longrightarrow \quad 
M' \propto {R'}^{(3\gamma -4)/(\gamma -2)}
\label{eq:LEmden}
\end{equation}
and for a polytrope of $\gamma = 5/3$ one finds $M' \propto {R'}^{-3}$,
exactly as in eq.~(\ref{eq:MR3rel}).

Now we can clearly see the advantages of solving the dimensionless TOV
equation. To get full dimensional results for a given fermion mass one
should simply rescale the dimensionless mass and radius with factors
$a=M_p^3/m_f^2$ and $b=M_p/m_f^2$, respectively. For the fermion mass
$m_f=1$~GeV, these factors are $a=1.632 M_\odot$ and $b=2.410$ km.
Therefore, for a point on the curve in Fig.~\ref{fig:MRrel} with the
dimensionless mass $ M'$ and the dimensionless radius $ R'$, the actual
mass and radius for a fermion star will be given by
\begin{equation}
M = 1.632 ~M_{\odot} \cdot M' \cdot \left(\frac{1 \mbox{
      GeV}}{m_{f}}\right)^2  
\label{eq:marbit}
\end{equation}
and 
\begin{equation}
R = 2.410 \mbox{ km } \cdot R' \cdot \left(\frac{1  \mbox{
      GeV}}{m_{f}}\right)^2 \quad .
\label{eq:rarbit}
\end{equation}
Accordingly, the maximum mass and minimum radius of a fermion star are
obtained for $M'=0.384$ and $R'_{min}=3.367$,
\begin{equation}
M_{max} = 0.627 ~M_{\odot} \cdot \left(\frac{1 \mbox{ GeV}}{m_{f}}\right)^2  
\label{eq:Mmaxacc}
\end{equation}
and 
\begin{equation}
R_{min} = 8.115 \mbox{ km } \cdot \left(\frac{1 \mbox{ GeV}}{m_{f}}\right)^2  
\label{eq:Rminacc}
\end{equation}
One can use the above relations to calculate the masses and radii of
different fermion stars. For example, a neutron star with $m_f=m_n =
939.6$ MeV in eq.~(\ref{eq:Mmaxacc}) and eq.~(\ref{eq:Rminacc}), has the
maximum mass $M_{max} = 0.710~M_{\odot}$ at a radius of $R_{min} =
9.192$ km, which match very well the original results of Oppenheimer and
Volkoff \cite{OV39}.

Next, we consider compact stars built of other fermions utilising the
curve shown in Fig.~\ref{fig:MRrel}. Besides the nucleon and electron,
fermions as neutrinos and quarks are well established in the standard
model. Moreover, other fermions such as neutralinos, axinos and
gravitinos are predicted in supersymmetric extensions of the standard
model. For ordinary neutrinos, the mass has been recently constrained to
the range from 50 meV to about 1 eV from the measurements of neutrino
oscillations and cosmological parameters (see \cite{SNO2004,WMAP2003}).
Hypothetical sterile neutrinos can have typical masses in the keV range
(see e.g.\ \cite{Dolgov2000}). For the supersymmetric particles,
neutralinos, gravitinos and axinos, there are large uncertainties
concerning their mass ranges. The most likely mass for neutralinos is
usually considered to be around 100 GeV but lighter neutralinos with a
mass of $\lesssim 50$ GeV are discussed in the literature, too
\cite{Bottino05}. Gravitinos can be heavy or very light in some models,
the latter mass range can extend from $10^{-2}$ eV to 1 keV, bounded by
constraints from big bang nucleosynthesis and the critical density (see
e.g.\ the discussion in \cite{Boerner_book}).  Axinos, the
supersymmetric partner of the axion, were introduced first as a possible
warm dark matter candidate with a mass in the keV range
\cite{Rajagopal91}, but they are discussed now also as a cold dark
matter candidate with much higher masses \cite{Covi99}.  Compact stars
could be also formed from composite dark matter candidates, similar to
white dwarfs, e.g.\ with the heavy charged stable leptons proposed in
\cite{Khlopov:2006uv}. Note, that the preceding discussion is just to
motivate interesting mass ranges for fermions to be considered in the
following. There are of course constraints for dark matter candidates
per se which are more severe than the ones mentioned above, see e.g.\
the recent review on dark matter candidates in \cite{Baltz04}.

\begin{table}
\begin{tabular}{c@{\hspace{2em}}cc@{\hspace{3em}}c}
\hline
Fermion mass & $M_{\rm max} (M_\odot)$ & $R_{\rm min}$ & comment
\\
\hline
100 GeV & $ 10^{-4}$ &  1 m  & neutralino star (cold dark matter) \\
1 GeV & 1 & 10 km & neutron star \\
1 GeV/0.5 MeV & 1 &  $10^3$ km & white dwarf \\
10 keV & $10^{10}$ & $10^{11}$ km & sterile neutrino star \\ 
1 keV & $10^{12}$ & $10^{13}$ km & axino star (warm dark matter) \\ 
1 eV & $10^{18}$ & $10^{19}$ km & neutrino star\\
$10^{-2}$ eV & $ 10^{22} $ & $ 10^{23} $ km & gravitino star \\
\hline
\end{tabular}
\caption{Maximum masses $M_{max}$ and radii $R_{min}$ for various cold
  compact stars made of a free Fermi gas}
\label{table:masses}
\end{table}

In principle, cold compact stars can be formed out of these exotic
fermions, too. Actually, compact stars made of massive neutrinos have
been introduced by Markov \cite{Markov64} and calculated within general
relativity by Gao and Ruffini \cite{Gao80}. Using the above mentioned
masses for these fermions, we have calculated their typical (i.e.\
maximum) masses and corresponding radii which are listed in
Table~\ref{table:masses}.  A compact star made of non-interacting
neutralinos has a maximum mass of about $10^{-4}$ solar masses with a
radius of about one meter. Warm dark matter, fermions with a mass in the
keV range, can form compact objects with galactic masses and a radius of
about one light year. In ref.~\cite{Bilic:2002hd,Bilic:2003ie} sterile
neutrinos with a mass of $m_f\sim 50$ keV were proposed to explain the
dark massive object at the center of our Galaxy, as an alternative to a
supermassive black hole. Compact stars made of dark matter interacting
with a scalar field have been also considered in
\cite{Tetradis:2005me,Brouzakis:2005cj}. Interestingly, if we take the
gravitino mass to be of the order of $10^{-2}$ eV, then the
corresponding gravitino star has the mass and the radius of our
universe. Note, that the above masses and radii are similar to the
free-streaming mass and radius scales which are of importance for
large-scale structure formation.

\subsection{Stars made of interacting fermions}

\subsubsection{Equation of State for Interacting Fermions}

The previous section was devoted to an idealised case of non-interacting
fermions. In a more realistic consideration, the inter-particle
interactions must be included, too. In the following, we address the
possible impact of interactions on the global properties of fermion
stars.

Consider the simplest model of two-body interactions between the
fermions.  In a lowest order approximation the interaction energy
density is proportional to $n^2$, where $n$ is the number density of
fermions. To have the correct dimensionality this term can be written as
$\rho_{int} = n^2/m_I^2$ where $m_I$ represents the energy scale of the
interaction. The corresponding contribution to the pressure is:
\begin{equation}
P_{int} = -\left.\frac{\partial E}{\partial V} \right|_{N,T=0} = n^2
\frac{(\partial \rho_{int}/n)}{\partial n} = \frac{n^2}{m_I^2} \quad .
\end{equation}
Therefore, in this approximation the energy density and pressure acquire
an additional term $n^2/m_I^2$.  The interaction must be repulsive so
that an increase in the number density increases the pressure and energy
density.  The scale $m_{I}$ can be also interpreted as the vacuum
expectation value of the Higgs field of the interaction. For weak
interactions, the interaction strength is just given by Fermi's constant
or the vacuum expectation of the Higgs field $v$ generating the masses
of the W and Z bosons, $G_F^2/\sqrt{2}=1/(2v^2)=(293 \mbox{ GeV})^{-2}$,
so that $m_I \sim 300$ GeV. Correspondingly, the strength of low energy
strong interactions, quantum chromodynamics (QCD), is controlled by the
pion decay constant $1/f_\pi^2$ in chiral perturbation theory, with
$f_\pi=92.4$ MeV being the vacuum expectation value of the sigma field
(for an introduction to chiral symmetry see e.g.\ \cite{Koch97}). In
quantum hadrodynamics, the expressions for the energy density and
pressure are exactly as given above for a repulsive interaction mediated
by a vector meson with an interaction strength of $g^2_{\omega
  N}/(2m^2_\omega)$ which is quite close to $1/f_\pi^2$ for $g_{\omega
  N} =13$ and $m_\omega=780$ MeV (see e.g.\ \cite{SW86}).  Hence, for
strong interactions the typical interaction mass scale is $m_{I}\sim
100$ MeV. In a more realistic approach, an attractive scalar interaction
should be included in addition to a repulsive vector interaction, too.
This kind of approach is widely used for baryonic matter \cite{SW86}
as well as for quark stars (see e.g.~\cite{Hanauske01}).  In
dimensionless variables the energy density and pressure can be written
as
\begin{eqnarray}
\rho' \equiv \frac{\rho}{m_f^4} &=& \frac{1}{\pi ^2} \int_0^{z}
x^2 \sqrt{1+x^2} dx  + \left(\frac{1}{3 \pi ^2}\right)^2 y^2 z^6
\nonumber \\ 
&=&  \frac{1}{8 \pi ^2} \left[ \left(2 z^3 + z\right)\left(1+
  z^2\right)^{\frac{1}{2}} - \sinh ^{-1}\left(z\right) \right] +
\left(\frac{1}{3 \pi ^2}\right)^2 y^2 z^6   
\label{eqi:den1}\\
p' \equiv \frac{p}{m_f^4} &=& \frac{1}{3 \pi^2} \int_0^{z}
\frac{x^4}{ \sqrt{1+x^2}}  dx + \left(\frac{1}{3 \pi ^2}\right)^2 y^2
z^6 \nonumber \\ 
&=& \frac{1}{24 \pi ^2} \left[ \left(2 z^3 - 3 z\right)\left(1+
    z^2\right)^{\frac{1}{2}} + 3  \sinh ^{-1}\left(z\right) \right] +
\left(\frac{1}{3 \pi ^2}\right)^2  
 y^2  z^6  
\label{eqi:pres1}
\end{eqnarray}
where $z$ is the dimensionless Fermi momentum and $y= m_{f}/m_{I}$ the
interaction strength.  

What are the values to be taken for the interaction strength $y$?  For a
realistic neutron star, neutrons interact strongly with $m_I \sim 100$
MeV, as outlined above. With a neutron mass of $m_f \sim 1$ GeV one
arrives at $y\sim 10 $. Neutrinos interact weakly and with a neutrino
mass of about 1 eV (for sterile neutrinos $m_{f}\sim 1$ keV), one finds
$y \sim 10^{-11} $ (for sterile neutrinos $y \sim 10^{-8}$). For
neutralinos with a mass of 100 GeV and weak interactions, $y \sim 1/3$,
for strongly interacting neutralinos $y \sim 10^3$.  We find that for
small values of $y$ the mass-radius relation remains almost unchanged
because the equation of state is dominated by the kinetic terms. Any
change in the equation of state can only occur for $y \geq 1$, i.e.\
when interaction term starts dominating the equation of state before the
fermions are becoming relativistic. Hence, it is sufficient to take $y$
to be in the range $10^{-2}$ to $10^3$. Below $y=10^{-2}$ one will
hardly observe any change in the mass-radius relation.

From the equations for $p$ and $\rho$ one finds that the common factor
$m_{f}^4$ can be taken out, so that general scaling arguments result in
$\epsilon_{\circ} = m_{f}^4$. Therefore, the dimensionless forms of $p$
and $\rho$ are $p'=p/m_{f}^4$ and $\rho'=\rho/m_{f}^4$ and the
corresponding mass $M$ and radius $R$ are $M'= M/a$ and $R'=R'/b$ with
$a = M_{p}^3/m_{f}^2$ and $b = M_{p}/m_{f}^2$, exactly Landau's mass and
radius as before.

Note, that for each value of $y$ there is a different equation of state
and two different regimes exist: $z \ll 1$, the non-relativistic limit,
and $z\gg 1$, the relativistic limit. For small $y\ll 1$, the equation
of state will be that of an ideal Fermi gas and for large $y\gg 1$ the
equation of state will be mostly determined by the interaction term,
unless $z$ becomes small enough so that the ideal gas term becomes the
dominant ones.

\begin{figure}
\begin{center} 
  \includegraphics[angle=-90,scale=.45]{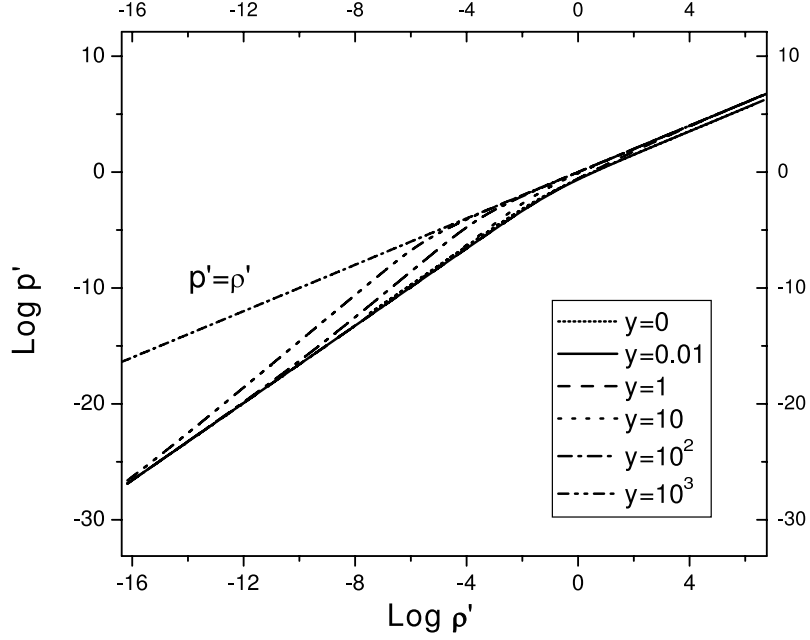}
  \caption{The equation of state in dimensionless form, $\log p'$ versus
    $\log \rho'$, for different values of the interaction strength as
    indicated in the figure.}
\label{fig:eqi}
\end{center}
\end{figure}

The dimensionless TOV equations (\ref{nd:tov1}) and (\ref{nd:tov2}) are
solved by making use of the dimensionless equation of state,
eqs.~(\ref{eqi:den1}) and (\ref{eqi:pres1}). The equation of state
depends on two parameters, $z$ and $y$. To solve the dimensionless TOV
equations for particular values of $y$, we first construct the
dimensionless equation of state in a tabular form for different values
of $z$.  Linear interpolation is used to find $p'$ corresponding to a
particular dimensionless density $\rho'$ and vice-versa. As before, the
dimensionless TOV eq.~(\ref{nd:tov1}) and eq.~(\ref{nd:tov2}) are
numerically solved using a fourth order Runge-Kutta algorithm. To plot
the graph of $M'$ versus $R'$ for various $y$, we take $100$ equally
spaced values of $y$ lying between $10^{-2}$ and $10^3$. For each value
of $y$, the $M'$ versus $R'$ curve is computed for $50$ equally spaced
dimensionless central densities. The program adjusts the step size in
the dimensionless radius to have about 2000 points for each star
configuration.  Fig.~\ref{fig:eqi} shows the final $ \log p'$ versus
$\log \rho'$ curve for a range of values of $z$ from $10^{-6}$ to $100$
for six different values of $y$.
  
From Fig.~\ref{fig:eqi} one makes the following observations:

\begin{enumerate}
\item For large values of $\rho'$, the different equations of state
  merge to one line, except for the curve of the cases $y=0$ and
  $y=0.01$ which stay slightly below the other ones.
\item The transition point from the relativistic curve
  to a steeper one shifts to lower densities for increasing
  interaction strength $y$.
\item For very small values of $\rho'$, the slope of the curve is
  approaching the slope for a non-interacting ideal gas. The higher the
  interaction strength $y$, the lower $\rho'$ must be taken to reach the
  non-interacting limit.
\end{enumerate}

These observations can be explained in the following manner. The
equations (\ref{eqi:den1}) and (\ref{eqi:pres1}) are parametric in
$z$ and $y$. In the non-relativistic limit, $z\ll 1$, they are reduced
to
\begin{equation}
\rho' \approx \frac{z^3}{3  \pi ^2} + \left(\frac{1}{3 \pi ^2}\right)^2
y^2  z^6  
\qquad \left(z\ll 1\right) 
\label{eq:zless1}
\end{equation} 
and
\begin{equation}
p' \approx \frac{z^5}{15  \pi ^2} + \left(\frac{1}{3 \pi ^2}\right)^2
y^2  z^6 
\qquad \left(z\ll 1\right) 
\label{eq:zless2}
\end{equation} 
In the relativistic limit, $z \gg 1$, one finds
\begin{equation}
\rho' \approx \frac{z^4}{4  \pi ^2} + \left(\frac{1}{3 \pi ^2}\right)^2  y^2
z^6  
\qquad \left(z\gg 1\right) 
\label{eq:zmore1}
\end{equation} 
and
\begin{equation}
p' \approx \frac{z^4}{12  \pi ^2} + \left(\frac{1}{3 \pi ^2}\right)^2
y^2  z^6 
\qquad \left(z\gg 1\right) 
\label{eq:zmore2}
\end{equation} 
If the interaction strength is small, $y\ll 1$, the interaction terms in
eqs.~(\ref{eq:zless1}) and (\ref{eq:zless2}) can be ignored for $z\ll 1$
and one recovers for small densities the non-interacting case. For
increasing $y$, the interaction terms become more and more important.
Let us examine the ratio of the interaction term to the free gas term in
more detail for the different cases.  We define the two ratios $
\tau_{\rho} (z\ll 1)$ and $ \tau_{p} (z\ll 1)$ for the dimensionless
energy density and pressure eqs.~(\ref{eq:zless1}) and (\ref{eq:zless2})
as
\begin{equation}
\tau_{\rho} (z\ll 1) \sim y^2  z^3 
\quad \mbox{ and } \quad
\tau_{p} (z\ll 1) \sim y^2  z
\end{equation}
The corresponding ratios $\tau_{\rho} (z\gg 1)$ and $\tau_{p}(z\gg 1)$
for the case $z \gg 1$ follow from eqs.~(\ref{eq:zmore1}) and
(\ref{eq:zmore2}),
\begin{equation}
\tau_{\rho} (z\gg 1) \sim y^2  z^2
\quad \mbox{ and } \quad
\tau_{p} (z\gg 1)\sim y^2  z^2 
\end{equation}
For non-relativistic fermions, $z \ll 1$, assume first that the
interaction strength $y$ is such that $\tau_{\rho} \ll 1$ and $\tau_{p}
\ll 1$. Then the interaction terms can be ignored and one recovers the
standard equation of state for a free non-relativistic gas, a polytrope
of $\gamma=5/3$. For larger values of the interaction strength, the
interaction terms become important already in the non-relativistic
regime, more drastically for the pressure, eq.~(\ref{eq:zless2}), than
for the energy density, eq.~(\ref{eq:zless1}), as for the same value of
$y$, $\tau_{\rho} \sim z^3$ while $\tau_{p} \sim z$. Hence, the curve
for the interacting case will be above the one for the non-interacting
gas, an effect which increases with increasing interaction strength $y$.
In the region where $\tau_{\rho} \gg 1$ and $\tau_{p} \gg 1$, the
interaction terms dominate and the equation of state reduces to $p' =
\rho'$, which is the stiffest equation of state consistent with special
relativity.  This equation of state was first considered by Zeldovich
\cite{Zeldovich61} (see the discussion in \cite{Zeldovich_book}).  Note,
that this condition is fulfilled already for $z\gg y^{-2/3}$, which for
large values of $y$ is substantially smaller than one. Therefore, even
for non-relativistic fermions one can reach the causal limit, where $p'
= \rho'$, for an strongly interacting gas at $\rho'\ll 1$.

For relativistic fermions, $z\gg 1$, consider first the interaction
strength $y$ to be such that $\tau_{\rho} \gg 1$ and $\tau_{p} \gg 1$.
Then the free gas terms in eqs.~(\ref{eq:zmore1}) and (\ref{eq:zmore2})
can be ignored and the equation of state reduces simply to $p' = \rho'$.
Thus, both in the relativistic and in the non-relativistic limit, the
equation of state will have the form $p' = \rho'$. For small values of
$y$, a different regime for the equation of state is reached when
$\tau_{\rho} \ll 1$ and $\tau_{p} \ll 1$. In that case the equation of
state becomes the one of an ultra-relativistic gas $p'=\rho'/3$, see
eq.~(\ref{ultra:rel}). This happens for e.g.\ $y = 0.01$ and $z \sim 10$
in Fig.~\ref{fig:eqi}.

With the above arguments we are now in the position to explain the
behaviour of the curves in Fig.~\ref{fig:eqi}. For $y=0.01$ and for
relativistic fermions, the equation of state is that for an
ultra-relativistic free gas, $p'=\rho'/3$, eq.~(\ref{ultra:rel}) which
differs from $p' = \rho'$, the one for an interaction-dominated Fermi
gas with large values of the interaction strength $y\gg 1$. The
different prefactor explains the slight difference between the lines for
$y=0.01$ and the other curves with larger values of $y$ for large
densities $\rho'$. There appears a sharp change of the slope of the
curves which signals the transition from the energy density being
dominated by interactions to being dominated by the (free) kinetic
terms.  For even smaller densities $\rho'\ll 1$, the equations of state
are eventually given by the non-relativistic polytrope of
eq.~(\ref{nonrel}).  Deviations from the non-interacting case arise for
intermediate densities with increasing interaction strength $y$, since
the interaction increases the pressure more rapidly than the energy
density.

\subsubsection{Solution to the TOV equation for interacting fermions}

\begin{figure}
\begin{center} 
  \includegraphics[angle=-90,scale=.45]{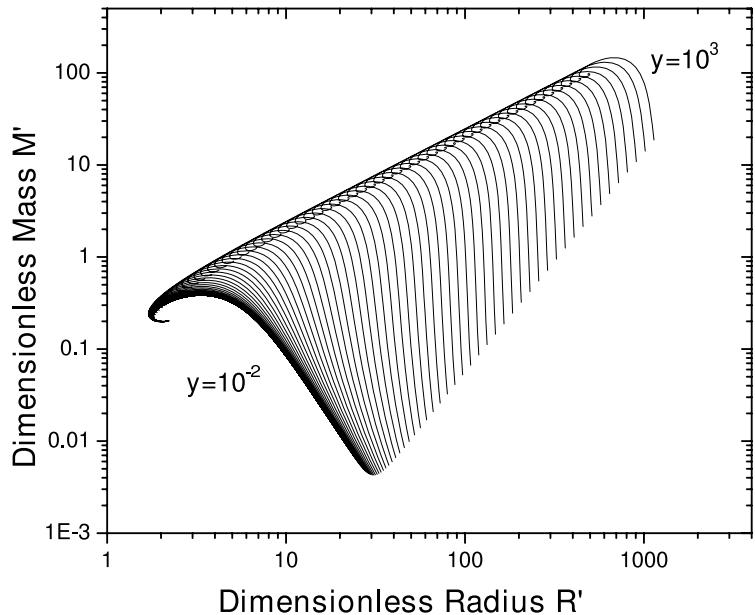}
  \caption{Mass-radius relation in dimensionless units for different
    values of the interaction strength $y$ in the range $10^{-2}$
    (lowest curve) to $10^3$ (highest curve) in 100 equal steps.}
\label{fig:mri}
\end{center}
\end{figure}

We solve the dimensionless TOV, eqs.~(\ref{nd:tov1}) and (\ref{nd:tov2}),
with the dimensionless pressure and energy density as given by
eqs.~(\ref{eqi:den1}) and (\ref{eqi:pres1}). The equation of state
depends now on the interaction strength parameter $y$. The plot of the
dimensionless mass $M'$ versus radius $R'$ for different values of $y$
is depicted in Fig.~\ref{fig:mri} on a double-log scale.
 
From Fig.~\ref{fig:mri} we observe that:

\begin{enumerate}
\item For small interactions strengths $y\ll 1$, the mass-radius curves
  are very close to each other. The interaction terms can be neglected
  in this case and the equation of state is determined by the free gas
  term, which results in nearly the same mass-radius curves.
\item For large interaction strengths, $y\geq 1$, the mass-radius curves
  are shifted towards larger masses and radii. As the interaction
  becomes stronger, the equation of state is getting more and more stiff
  approaching $p'=\rho'$.
\item The mass-radius curves, in particular the maximum mass and the
  corresponding radius, seem to follow a common trend for strongly
  interacting fermions, so a scaling behaviour should be expected.
\item The slope of the mass-radius curve for masses well below the
  maximum mass changes with increasing interaction strength $y$ from the
  free case behaviour $M' \times {R'}^3 = const.$ to one with a rather
  constant radius for a wide range of masses.
\end{enumerate}

In order to obtain the dimensionful mass and radius for different values
of $y$ one should use eqs.~(\ref{eq:marbit}) and (\ref{eq:rarbit}) and
take the numerical values for $M'$ and $R'$ from the corresponding curve
in Fig.~\ref{fig:mri}. For a neutron star, strong interactions between
neutrons mediated by vector mesons sets the typical interaction strength
to be around $y=10$.  Then, the maximum stable mass emerges to be $3.77
M_{\odot}$ instead of $0.71 M_{\odot}$ as was found earlier for
non-interacting neutrons. Note, that attractive forces will reduce this
value somewhat closer to the maximum masses considered presently for
realistic equations of state (see e.g.\ \cite{Lattimer2001}). For an
interacting neutralino star, weak interactions result in low values for
the interaction strength, $y\sim 1/3$, so that the maximum mass would
stay unchanged. However, if we assume strong interaction values for $y$
as motivated by QCD of around $y =10^3$, the maximum stable mass is
boosted to $2.7 \times 10^{-2} M_{\odot}$ compared to $6.3 \times
10^{-5} M_{\odot}$ for the non-interacting case.

Next we discuss the behaviour of the mass-radius relation for large
values of $R'$ considering strongly interacting fermions, i.e.\ $y\gg
1$. The question is, is there any relation similar to the one for the
non-interacting case, eq.~(\ref{eq:MR3rel})? Note first, that the tail
of the $M'$ versus $R'$ curve corresponds to small central densities of
the star $\rho'(0)$, so that the dimensionless Fermi momentum
$z=k_f/m_f$ is quite small, i.e.\ $z \ll 1$. In the
non-relativistic regime, it is likely that the equation of state can be
described by a polytrope, even for interacting fermions. 

Indeed, at $z \ll 1$, the equation of state reduces to
eqs.~(\ref{eq:zless1}) and (\ref{eq:zless2}) and one can ignore the
interaction term proportional to $z^6$ as compared to the 
term proportional to $z^3$.  Thus one finds that
\begin{equation}
\rho' \simeq \frac{z^3}{3  \pi ^2}
\label{eq:z3term}
\end{equation} 
Now, using this equation to eliminate $z$ from eq.~(\ref{eq:zless2}), we
get
\begin{equation}
p' \simeq \frac{(3 \pi ^2)^{2/3}}{5} {\rho'} ^ {5/3} + y^2  {\rho'}^2
\label{eq:simplify_p}
\end{equation} 
For large values of $y\gg 1$, the interaction term becomes important for
the pressure above a certain Fermi momentum, when $\tau_p(z\ll 1) \sim
1$ or $z\sim 1/y^2$.  Then, the pressure grows with the square of the
energy density and the corresponding polytropic coefficient is $\gamma =
2$.  Using the Lane-Emden solution, (\ref{eq:LEmden}), one finds that
the radius does not depend on the central density, while the mass
increases linearly with the central density, hence $R'=const.$ for a
large range of masses. Examining Fig.~\ref{fig:eqi}, we see that this
behaviour is present for $y\gg 1$ close to the point where the causal
limit, $p'=\rho'$, is reached.  The upper limit to the energy density is
therefore given by the point where the interaction starts to affect the
energy density in addition to the pressure, i.e.\ $\tau_\rho (z\ll
1)\sim 1$ or $\rho\sim z^3\sim y^{-2}$. The critical lower value is
given as discussed above by the condition $\tau_p (z\ll 1)\sim 1$, i.e.\
$z\sim y^{-2}$ or $\rho'\propto y^{-6}$, where the interaction starts to
affect the pressure. Therefore, radii are constant for the mass range
extending from $M'\propto \rho'\propto y^{-6}$ to $M'\propto\rho'\propto
y^{-2}$, which can reach, for the case $y=10^3$, up to about twelve
orders of magnitude in compact star masses!

\subsubsection{Scaling behaviour for interacting fermions}

As mentioned above, a scaling behaviour is expected for large values of
$y$. Consider eq.~(\ref{eq:zless1}) and eq.~(\ref{eq:zless2}) for
non-relativistic fermions and for $y \gg 1$. Then the dimensionless
pressure $p'$ is dominated by the interaction term while the
dimensionless energy density $\rho'$, (\ref{eq:zless1}), contains both
the kinetic and interaction terms. One can see that in this case the
equation of state will turn from the non-relativistic polytrope,
(\ref{nonrel}), to the causal limit of the form $ p'= \rho'$ when the
fermions are still non-relativistic, i.e.\ $z \ll 1$. Indeed, one
can write the dimensionless energy density as:
\begin{eqnarray}
\rho' &=& \frac{z^3}{3  \pi ^2} + (\frac{1}{3 \pi ^2})^2  y^2  z^6 
\nonumber \\
&=& \frac{z^3}{3  \pi ^2} + p'
\label{eq:compare}
\end{eqnarray}
Hence, the pressure $p'$ is larger than the kinetic term of the energy
density for $z^3 \sim y^2 z^6$ or $z^3 \sim 1/y^2$. In this regime, both
$p'$ and $\rho'$ are of the order of $\sim 1/y^2$. Hence, the pressure
and energy density can be rescaled in a dimensionless form with the
factor $m_f ^4/y^2$. The corresponding Landau mass and Landau radius
will be modified accordingly to 
\begin{equation}
M_L^{int} = M_p^3/m_f^2 \cdot y \quad \mbox{ and } \quad 
R_L^{int} = M_p/m_f^2 \cdot y \quad,
\end{equation}
respectively, so that the maximum mass and the corresponding radius
increase linearly with the interaction strength $y$. We note in passing
that the importance of the interaction on the global properties of
compact objects has been also noted for the case of boson stars with
interacting scalar fields in \cite{Colpi86}.

\begin{figure}
\begin{center} 
  \includegraphics[angle=-90,scale=.45]{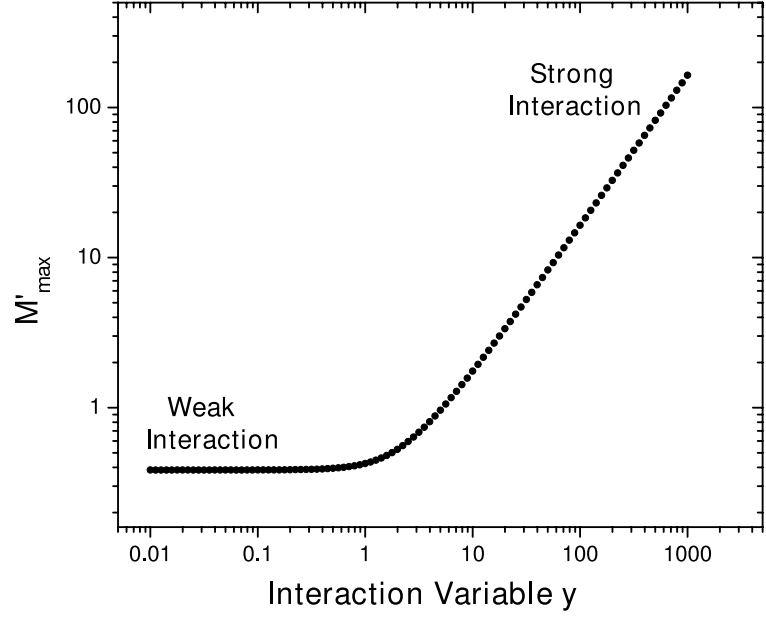}
  \caption{The dimensionless maximum mass $M'_{max}$ for interacting
    fermions versus the interaction strength $y$ on a double-log scale.
    For small values of $y$, the maximum mass does not change, while it
    increases with a power law for strong interactions ($y > 1$).}
\label{fig:mvsy}
\end{center}
\end{figure}

The numerical results for the scaling behaviour of the maximum mass as a
function of the interaction strength $y$ are plotted in
Fig.~\ref{fig:mvsy} on a double-log scale.  For small values of $y$, the
maximum mass $M'_{max}$ does not change and basically remains constant.
For $y \sim 1$, the maximum mass $M'_{max}$ starts increasing as a
function of $y$, approaching a power-law rise for $y \gg 1$.

For small values of $y$, the interactions do not affect significantly
the equation of state and there will be almost no change in the maximum
mass. If the interaction terms in the equation of state become
comparable to the ones of the free gas, i.e.\ if $y\sim 1$, the maximum
mass begins to increase.  Finally, for $y \gg 1$, the pressure and the
energy density for the maximum mass configuration are dominated by the
interaction terms, which can be scaled out by the factor $1/y^2$ so that
the maximum mass will increase linearly with $y$ as explained above. For
the part of the graph in Fig.~\ref{fig:mvsy} where $y\gg 1$, one can
perform the following general fit:
\begin{equation}
M'_{max} = c_1 + s_1 \cdot y^{\gamma_1} \quad ,
\end{equation}
as $ \log M'_{max}$ is linear in $\log y$ for $ y \gg 1$. The constant
$c_1$ is fixed by the numerical result found for the non-interacting
case, hence $c=0.384$. The parameters $s_1$ and $\gamma_1$ are fitted to
the curve of Fig.~\ref{fig:mvsy} to be $s_1=0.165$ and $\gamma_1 = 0.999
\approx 1$. The maximum mass $M_{max}$ of a compact star can then be
calculated from the relation
\begin{equation}
M_{\rm max} = \left(0.384 + 0.165 \cdot y \right)\cdot \left(\frac{1~\rm
    GeV}{m_{f}}\right)^2 \cdot 
1.632  M_{\odot} 
\label{eq:mvsy}
\end{equation}
as an approximation to our numerical solution of the full TOV equations.

\begin{figure}
\begin{center} 
  \includegraphics[angle=-90,scale=.45]{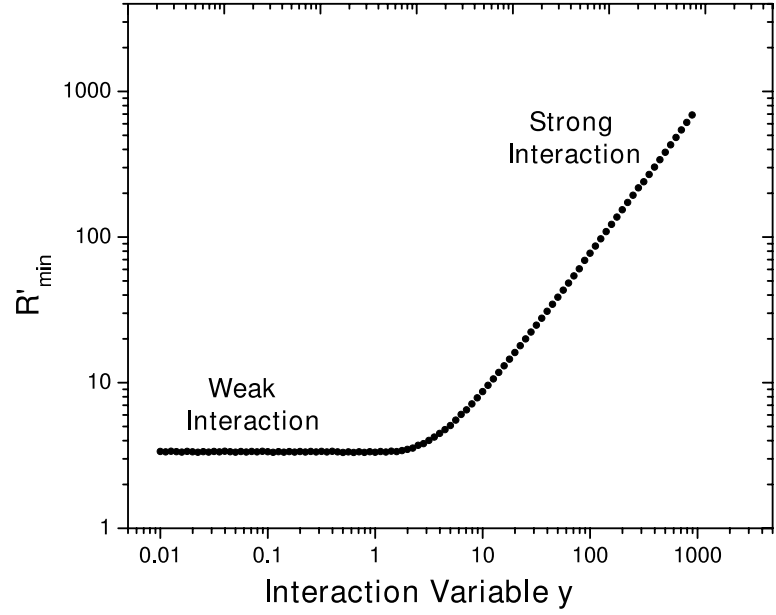}
\caption{The minimum dimensionless radius $R'_{min}$ for the maximum
  mass configurations as a function of the interaction strength $y$ on a
  double-log scale.}
\label{fig:rvsy}
\end{center}
\end{figure}

We plot the minimum radius $R'_{min}$ versus $y$ on a double-log scale
in order to extract a scaling behaviour for the radius in
Fig.~\ref{fig:rvsy}. For small values of $y$, the radius $R'_{min}$
stays constant. For $y>1$, the radius increases with the interaction
strength following again a power-law behaviour, so that we take the
following fitting expression for $R'_{min}$:
\begin{equation}
R'_{min} = c_2 + s_2 \cdot y ^ {\gamma_2}
\end{equation}
where $c_2$, $s_2$ and $\gamma_2$ are constants. The constant $c_2$ is
fixed by the non-interacting limit as discussed before, i.e.\ by
$R'_{min}$ for a free Fermi gas, for which numerically $c_2 = 3.367$.
In the $y\gg 1$ limit, one obtains $s_2=0.797$ and $\gamma_2 =
0.9942\sim 1$. Hence, for large interactions strengths $ y\gg 1 $ the
radius can be approximated as
\begin{equation}
R_{min} = \left( 3.367 + 0.797 \cdot y \right) \cdot 
\left(\frac{1~\rm GeV}{m_f}\right)^2 \cdot 2.410  \mbox{ km}
\label{eq:rvsy}
\end{equation}
According to the scaling arguments as derived above, the minimum radius
(as well as the maximum mass) should increase linearly with $y$ for
large interaction strengths which is indeed being found numerically and
observed in Fig.~\ref{fig:mvsy} for the maximum mass and in
Fig.~\ref{fig:rvsy} for the corresponding radius. For small values of
$y$, scaling arguments predict a nearly constant minimum radius as a
function of $y$ as also clearly seen in Fig.~\ref{fig:mvsy} and in
Fig.~\ref{fig:rvsy}.

\begin{figure}
\begin{center} 
  \includegraphics[angle=-90,scale=.45]{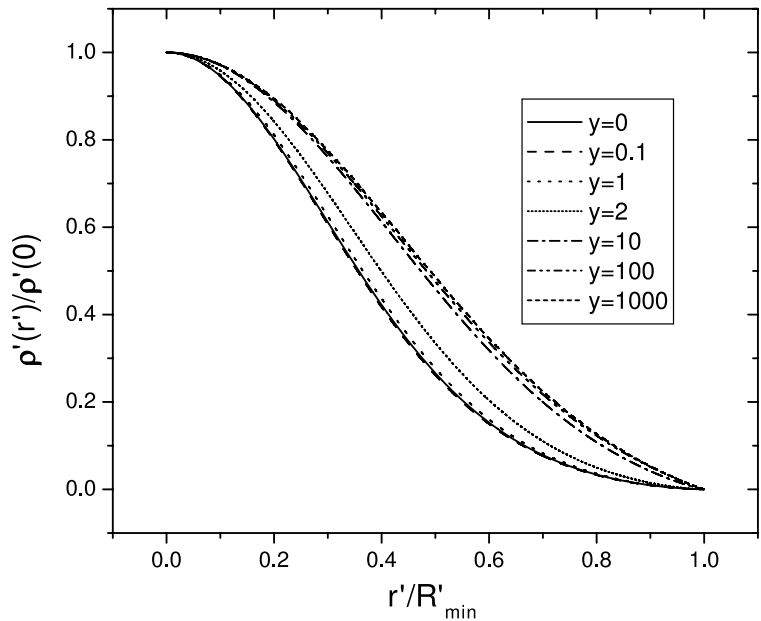}
\caption{The energy density $\rho'(r')$ versus the
  radius $r'$, for star configurations with different values of
  the interaction strength $y$.}
\label{fig:core}
\end{center}
\end{figure} 

However, one should expect a difference in the radial density profile
for the weak and the strong interaction cases. When interactions are
weak, the equation of state is governed by the ideal Fermi gas terms in
the core of the compact star.  For increasing interaction strength, the
pressure increases more rapidly due to the interaction terms for low
energy densities. As a result, at the surface of the compact star, the
energy density decreases more rapidly and the additional pressure from
the interactions becomes important.  A halo is created as more matter
can be supported against the gravitational pull. The total mass of the
star remains almost the same as it is mainly given by the dense core and
not affected by the dilute halo. Interestingly, also the corresponding
radius is not modified substantially by this effect. This behaviour is
illustrated in Fig.~\ref{fig:core} which shows the density profile for
various star configurations with different values of the interaction
strength $y$.  The dimensionless density $\rho'(r')$ is plotted versus
the dimensionless radius of the star $r'$ for several configurations.
Note, that the curves are drawn for the maximum mass configurations and
normalised to their central density and the total radius.
Fig.~\ref{fig:core} shows the presence of a dense core and a low density
halo at the surface of the star. The rate of the decrease of $\rho'(r')$
with the radius $r'$ is large for small values of $y$ and decreases
gradually for higher values of $y$ generating a denser halo of the
compact star for strongly interacting fermions compared to the weakly or
non-interacting ones.

\section{Summary and discussions}
\label{sec:summary}

We construct general equations of state for fermions of arbitrary mass
$m_f$ to be used as input for solving the TOV equations for
corresponding compact stars.  Besides a free gas of fermions, we
consider also the case of interacting fermions by adding interaction
terms $\sim n^2$ to the energy density and pressure. We discuss various ways of
rescaling the TOV equation and find the corresponding scaling solutions
for compact star configurations for arbitrary fermion mass and
interaction strength. The scaling solutions are tested by detailed
numerical calculations.

In particular, we have demonstrated that Landau's argument and the
corresponding expressions for the maximum mass and the corresponding
(minimum) radius hold also in the case of general relativity.  For a
compact star made of fermions, there exists an upper limit for the mass
which is of the order of the Landau mass $M_p^3/m_f^2$, where $M_p$ is
the Planck mass. The corresponding radius scales as the Landau radius
$M_p/m_f^2$. By directly solving the TOV equations, we have found the
mass-radius relation in dimensionless form, i.e.\ in units of the Landau
mass and Landau radius, supporting our analytic finding. The numerical
values of the maximum mass and the corresponding radius are $M_{max} =
0.384 M_p^3/m_f^2$ and $R_{min} = 3.367 M_p/m_f^2$, respectively. For
large radii the mass and the radius are related by the well-known
relation $M \cdot R^3 = const$. The results are in accordance with those
known for a free neutron gas but our scaling relations can be used for
fermions of any mass.  For example, fermions with a mass of 100 GeV can
form compact objects up to a maximum mass of about $10^{-4} M_\odot$ and
radii down to 1 meter as deduced from the same mass-radius relation as
for a free gas of neutrons, when properly rescaled by the Landau mass
and Landau radius.  The same can be done for the mass-radius relation of
neutrino stars.  For $m_f\simeq 1$~eV the corresponding maximum mass and
its radius are, however, of the order of $M_{max}\sim 10^{18} M_\odot$
and $R_{min}\sim 10^{19}$ km.  Interestingly, a hypothetical fermion
with a mass of about $10^{-2}$ eV can have a maximum mass and a
corresponding radius which matches the total mass and the horizon of the
present universe.

Effects from interactions between the fermions are taken into account by
adding terms proportional to the fermion density squared to the
expressions for the pressure and energy density. This can be motivated
by e.g.\ such effective models of strong interactions as quantum
hadrodynamics. The resulting equation of state depends on a new variable
$y = m_{f}/m_{I}$ which controls the strength of the interaction. The
mass $m_I$ fixes the range and strength of the interactions. The values
of $y$ can be as small as $10^{-11}$ for neutrinos with a mass of
$\sim 1$~eV and as high as $y=10^3$ for strongly interacting
neutralinos with a mass of $\sim 100$~GeV.

Using scaling arguments we arrive at the dimensionless equation of state
and corresponding scaling solutions for the maximum mass and its radius
for interacting fermions. We show that the maximum mass and the radius
are rather constant for small interaction strengths, $y\ll 1$, but are
entirely determined by the interaction terms for $y\gg 1$: Both, the
maximum mass and its radius increase linearly with the interaction
strength for $y\gg 1$. The scaling is supported by numerical
calculations, where we find that $M_{max} \approx 0.165 y \cdot
M_p^3/m_f^2$ and $R_{max} \approx 0.797 y \cdot M_p/m_f^2$ for $y\gg 1$.
Besides these general scaling features, there is a small change in the
density profile for large interaction strengths, which can be attributed
to the formation of an enhanced halo density in the outer region of the
star. For large interaction strengths the mass-radius relation changes
from the standard $M\cdot R^3 = const$ behaviour to the one with a
constant radius for a wide range of compact star masses. This is because
the pressure increases as the square of the energy density and not as
the power 5/3 for a free non-relativistic Fermi gas. The mass range for
constant radii changes with the interaction strength in the range
between $y^{-6}$ and $y^{-2}$. Specifically, for values of the
interaction strength of $y=10^3$ and a fermion mass of 100 GeV, the
maximum mass increases from the value for the non-interacting case,
$M_{max} \sim 10^{-4} M_\odot$ to $M_{max} \sim 10^{-1} M_\odot$, which
is comparable to the one for ordinary neutron stars.  The compact star
mass range, where the radius stays constant, extends from about the
maximum mass of about $10^{-1} M_\odot$ down to $10^{-13} M_\odot$ with
a typical radius as given by the Landau radius, i.e.\ about 100 meter.

There are two important issues which require a special study which is
beyond the scope of the present paper.

First, the crucial assumptions for all these investigations is that the
fermions are stable on the time scale comparable with the lifetime of
the universe, i.e.\ $\tau\geq H_0^ {-1} \approx 14$ Gyr, where $H_0$ is
the present value of the Hubble constant. In other words, it is assumed
that the fermions constituting the compact star are conserved, i.e.\
there is no annihilation into other kinds of matter.  Naive estimates
for the lifetime using $\tau \sim (n\cdot \sigma)^{-1}$, with the number
density $n\sim m_f^3$ and the cross section $\sigma \sim m_f^2/m_I^4$,
result in the constraint $m_f < (H_0 \cdot m_I^4)^{1/5}$.  For weak
interactions, one arrives at fermion masses $m_f < 1$ keV, for
gravitational interactions at the Planck scale $m_I=M_p$, however, the
fermion mass must be only lower than $m_f < 10^4$~TeV (in all cases
$y\ll 1$).

The second important question is, when and how the compact objects made
of exotic fermions could be formed? One may speculate that this could
happen at very early stages in the history of the universe, right after
the inflation stage. These early formed objects could serve as seeds for
clumping ordinary matter at later stages of the expansion, after the
radiation decoupling. Therefore, one may speculate about hybrid objects
where exotic fermion stars are surrounded by the halos of ordinary
matter.

Our final remark concerns the possible observations of compact objects
made of dark matter particles. In fact, there exist limits derived from
the observation of micro-lensing events. The MACHO collaboration has
excluded the mass range of $(10^{-7} - 30) M_\odot$ for compact objects
forming the bulk of the Galactic dark matter. However, compact stars of
these mass ranges are not ruled out if they do not contribute more than
$4\cdot 10^{11} M_\odot$ to the Galactic halo (see \cite{Alcock2001} and
references therein).

\begin{acknowledgments}
  This work has started as a summer research project of Gaurav Narain at
  the Goethe University in Frankfurt, Germany. GN thanks the institute
  for theoretical physics for their warm hospitality. JSB thanks Stefan
  Hofmann for several discussions on dark mater and dark stars. This
  work is supported in part by the Gesellschaft f\"ur
  Schwerionenforschung (Germany), and the grants RFBR 05-0204013 and
  NS-8756.2006.2 (Russia).
\end{acknowledgments}
 
\bibliography{all,literat}

\end{document}